\documentclass[runningheads]{llncs}
\usepackage[T1]{fontenc}
\usepackage{graphicx}
\usepackage{amsmath}
\begin{document}
\title{Learning Melanocytic Cell Masks from Adjacent Stained Tissue\thanks{Medical Image Learning with Limited \& Noisy Data Workshop at MICCAI 2022.}}
%
%
\author{
Anonymous
}
\author{
Mikio Tada\inst{1}
\and
Ursula E. Lang\inst{1}
\and
Iwei Yeh\inst{1}
\and
Elizabeth S. Keiser\inst{1}
\and
Maria L. Wei\inst{1,2}
\and
Michael J. Keiser\inst{1}
}
\authorrunning{M. Tada et al.}
%
\institute{University of California, San Francisco, CA, USA 
\and San Francisco VA Health Care System, San Francisco, CA, USA}
\maketitle              
\begin{abstract}
Melanoma is one of the most aggressive forms of skin cancer, causing a large proportion of skin cancer deaths. However, melanoma diagnoses by pathologists shows low interrater reliability. As melanoma is a cancer of the melanocyte, there is a clear need to develop a melanocytic cell segmentation tool that is agnostic to pathologist variability and automates pixel-level annotation. Gigapixel-level pathologist labeling, however, is impractical. Herein, we propose a means to train deep neural networks for melanocytic cell segmentation from hematoxylin and eosin (H\&E) stained sections and paired immunohistochemistry (IHC) of adjacent tissue sections, achieving a mean IOU of 0.64 despite imperfect ground-truth labels.

\keywords{Pathology  \and Deep learning \and Image segmentations.}

\end{abstract}

\section{Introduction}
Melanoma comprises only 1\% of all skin cancer types, yet causes 97\% of deaths due to skin cancer~\cite{Siegel2022}. The five-year survival rate for melanoma is 99\% when diagnosed at an early stage. However, it drops to 68\% when diagnosed after regional spread and to 30\% when diagnosed after distal metastases have occurred~\cite{Siegel2022}. Early diagnosis and confirmation of full surgical tumor excision is crucial to decrease the risk of death. Melanoma develops from pigment-producing melanocytes. In the process of melanoma diagnosis, pathologists may employ IHC stains that use antibodies targeting melanocyte markers to confirm melanocytic lineage and determine the extent of the tumor. However, IHC stains are costly and time consuming to process. The objective of our study is to develop an automated annotation method that segments melanocytic cells at the pixel-level and acts as an in silico IHC stain, inferred solely from standard H\&E stains.

\section{Related Studies}
Segmentation labels for dermatopathology whole slide images (WSIs) pose three major issues. (1) Manual pixel-level labeling is exhausting and impractical at the gigapixel level, especially when the objects annotated, such as individual and at times sparsely-spaced cells, are fine-grained. (2) Annotation requires domain expertise. Several prior studies nonetheless obtained melanocyte labels based on pathologists or dermatopathologists’ manual annotations~\cite{Lu2013,Kucharski2020,Liu2021}, but even these were restricted to limited regions of interest. (3) Low interrater reliability even among trained experts~\cite{Elmore2018} limits generalizability of models trained on labels, if not from a representative expert cohort, compounding the challenge. In pathology, IHC stains can highlight specific cell types or structures, providing an alternative, data-driven label, as in studies involving segmentation of epithelium~\cite{Bulten2019}, mitosis detection~\cite{Tellez2018}, and melanocyte nuclei~\cite{Jackson2020}. Our study similarly uses IHC stains to label melanocytes, however, it does not rely on destaining techniques~\cite{Jackson2020} that may damage tissue and are not routinely available in clinical settings.

\section{Method}
\subsection{Dataset}
Our dataset consists of 22 pairs of H\&E and IHC stained WSIs scanned by Aperio AT2. For the IHC, we used the Melan-A antibody, specific for melanocytic cells, which stains cell cytoplasm. Each pair of H\&E and IHC slides were cut adjacently, to maximize morphological similarity, unlike~\cite{Bulten2019,Tellez2018,Tellez2018} where the tissues were stained with H\&E, destained, then restained using IHC. All 22 pairs of slides were manually labeled at the WSI-level by two dermatopathologists to one of four different stages of tumor progression - benign nevus: 7 slides, dysplastic nevus: 8, melanoma in situ (MIS): 3, and invasive melanoma: 4.

\subsection{Automated melanocytic cell label generation}
\begin{figure}
    \centering
    \includegraphics[width=\textwidth]{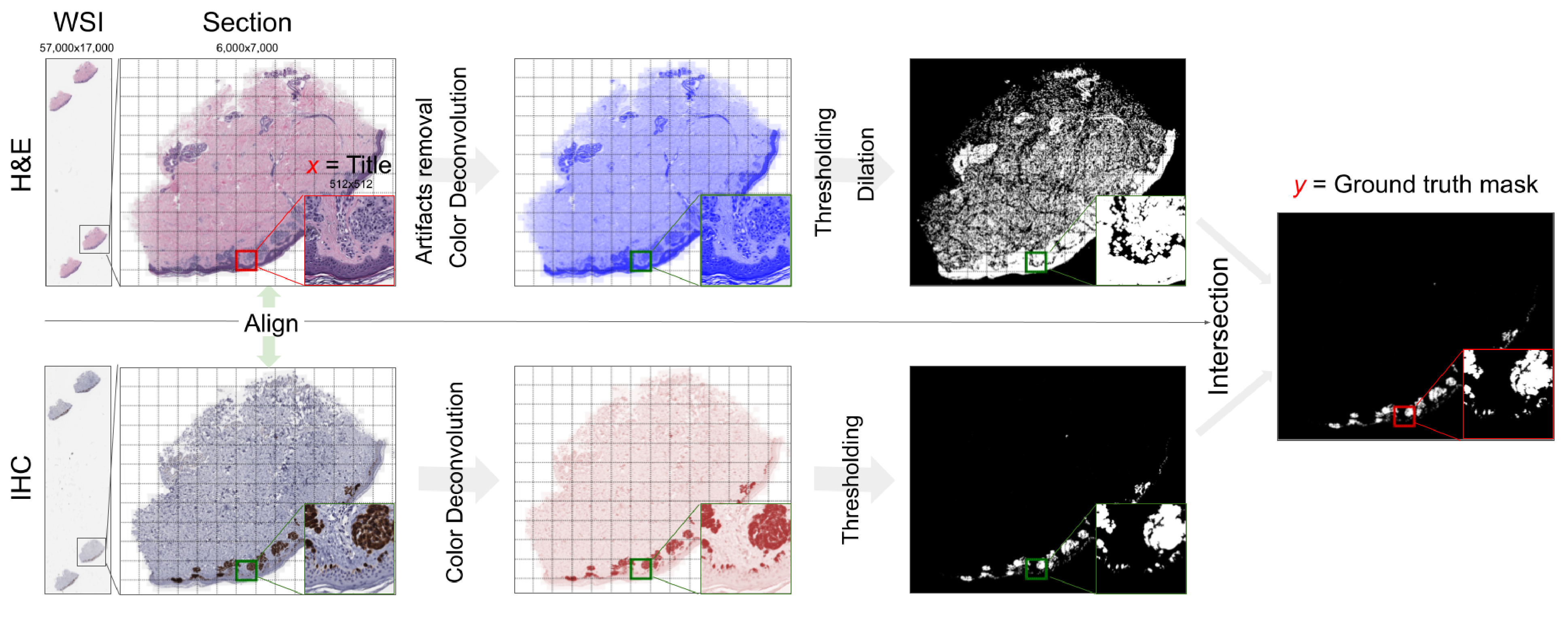}
    \caption{Overview of an automated method to generate ground truth label masks using H\&E and IHC stained tissues. The workflow begins (at left) with H\&E and IHC WSIs and ends with a binary melanocyte ground-truth mask. The operations are done at the section-level, the model takes 512x512 tiles as input.} \label{Figures1}
\end{figure}

\noindent
For each matched pair of H\&E and IHC WSIs, we generated melanocytic cell label masks from the IHC (Figure 1). The pipeline had to overcome noisy-data limitations in several steps:
\begin{enumerate}
  \item We aligned individual H\&E tissue sections with the corresponding IHC section. Fig. 2 (a) top shows a misaligned pair; Fig. 2 (a) bottom shows sections registered with ImageJ’s bUnwarpJ~\cite{imagej}. 
  \item We removed artifacts such as spurious blue ink by manual color range Fig. 2 (b).
  \item We separated hematoxylin from the H\&E image and diaminobenzidine from the IHC image using color deconvolution from the HistomicsTK Python library~\cite{histomicstk}.
  \item We thresholded nuclei and melanocytic cell images by the color histograms of nuclei and melanocytic cells. Our study used a cutoff value of 60.
  \item Because Melan-A IHC is cytoplasmic, we dilated the nuclei segmentation from step 4. For parsimony and with assistance from a dermatologist, we dilated the initial nuclei 10 times, Fig. 2 (c).
  \item We intersected IHC melanocytic cells masks with cytoplasm masks from H\&E, to avoid false-positive label pixels due to cellular misalignment or tissue z-axis differences.
  \item We tiled the sections into 512x512 pixel input H\&E images, with the masks as the labels.
\end{enumerate}

\subsection{Model training and evaluation}
A total of 22 slides were split into 9 slides for training (3 benign, 4 dysplastic, 1 MIS, 1 melanoma), 6 for validation (1 benign, 2 dysplastic, 1 MIS, 2 melanoma), and 7 for test (3 benign, 2 dysplastic, 1 MIS, 1 melanoma). The ratio of melanocytic cell-positive pixels for each of four diagnostic classes was approximately 70:15:15 for training, validation, and test sets.
We trained UNet~\cite{Ronneberger2015} with AdamW optimizer~\cite{Loshchilov2017}, with an initial learning rate of 3e-4. The loss function was a simple sum of Focal loss~\cite{Lin2017} and Tversky loss~\cite{Salehi2017}. The training ended after the validation loss stopped decreasing for 8 epochs. Evaluation metric was mean Intersection Over Union (IOU), a commonly used pixel-level metric:

\begin{equation}
    \small 
    \text{mean IOU} =  \frac{1}{2} \times \left(\frac{\text{TP}}{\text{TP}+\text{FP}+\text{FN}} + \frac{\text{TN}}{\text{TN}+\text{FP}+\text{FN}}\right)
\end{equation}

\section{Results}
With the intention to automate annotations of melanocytic cells, our method (Fig. 1), instead of relying on human manual annotations~\cite{Jackson2020}, only consists of medical image processing techniques such as image registration, color deconvolution, color thresholding, and dilation. These techniques enabled us to minimize erroneous ground truth masks of melanocytic cells and limited the inherent noisiness arising from H\&E and IHC staining procedures. Aligning H\&E and IHC tissues was a crucial step (Fig. 2 (a)) because even a slight shift between large image alignments resulted in false masking of melanocytic cells at inaccurate areas of the H\&E stained tissue. Additionally, we had to remove spurious blue surgical ink arising from the slide processing (Fig. 2 (b)). Otherwise, color deconvolution falsely identified those artifacts as nuclei, resulting in false-positive melanocytic cell masks. We discarded 10\% of the initially collected tissue sections due to too many false-positives of melanocytic cell masks. Because shapes and sizes of cells can vary partially due to different melanoma stages and also due to the cutting of cells at different cross sections, we had to dilate the initial nucleus segmentation 10 times (Fig. 2 (c) right). Lastly, we found that intersecting IHC melanocytic cells masks with H\&E cytoplasmic masks generated more stable masks than simply projecting IHC highlighted melanocytic cells onto the H\&E stained sections without correction. A model trained on the intersection of melanocytic cell and cytoplasmic masks outperformed the projection trained model, based on the intersection benchmark as well as IHC alone benchmark. 

\begin{figure}
    \centering
    \includegraphics[width=\textwidth]{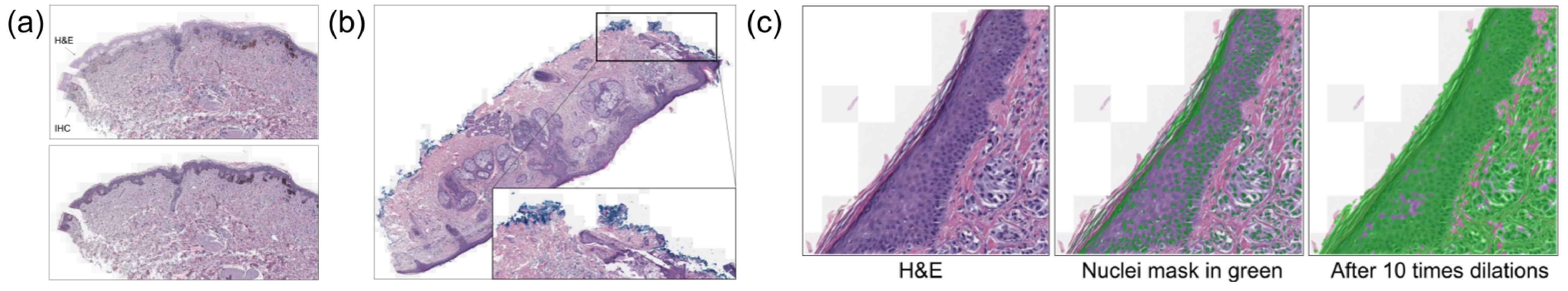} 
    \caption{(a) Effect of image registration to alignment between H\&E and IHC. (b) An example of spurious blue ink. (c) Demonstration of expanding the initial nuclei segmentation to cytoplasm. } \label{Figures2}
\end{figure}

\noindent
We observed that the model expressed high sensitivity to melanocytic cells but tended to also include the surrounding dermis in its positive predictions. At a prediction threshold of 0.5, Fig 3. (a) shows a few areas of false negatives (in red), yet a considerable amount of false positives (orange), especially for benign nevus. Tissue sections with melanoma achieved notably higher mean IOUs (Fig. 3 (b)) than benign nevus, likely because melanoma has more absolute volume of melanocytic cells. Overall, the model was able to localize general locations of melanocytic cells, as Fig. 3 (b) showing above mean IOU 0.64 for most of the sections. We cannot directly compare this method of creating melanocytic cell ground truth masks or our segmentation model to previous studies because neither their code nor datasets are publicly available.

\begin{figure}
    \centering
    \includegraphics[width=\textwidth]{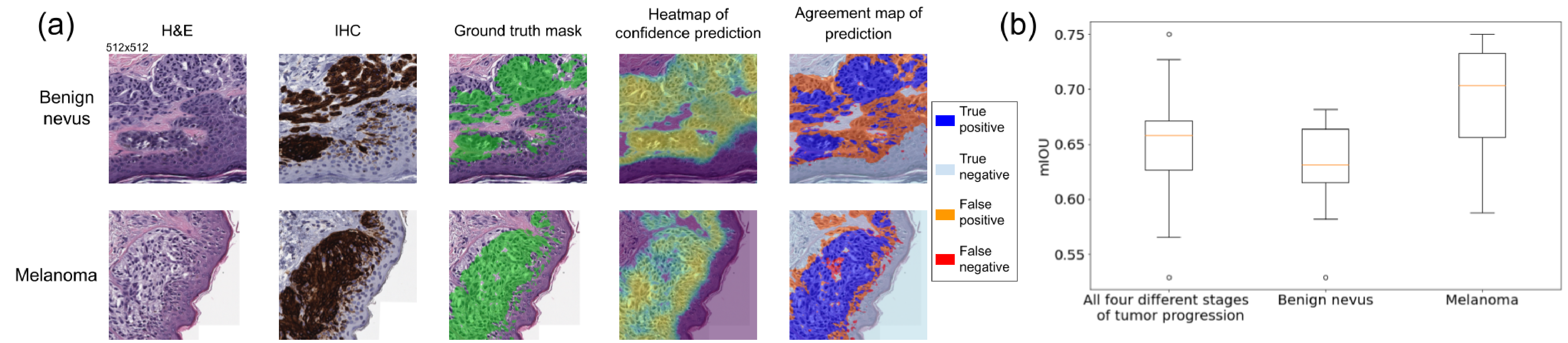}
    \caption{(a) From left to right, examples of pairs of H\&E and IHC tiles, the ground truth (green), the raw prediction, and the agreement between ground truth and the class prediction, for benign nevus and melanoma. (b) Distribution of mean IOUs based on the entire test set, benign nevus alone, and melanoma alone.} \label{Figures3}
\end{figure}

\section{Conclusion}
Our study demonstrates a means to automatically annotate melanocytic cells using paired H\&E and Melan-A IHC tissue sections and a deep neural network that can segment melanocytic cells in H\&E stained tissues. We envision that our proposed method could be more widely used with other tissue types and antibodies to automate the labeling of pathology images at the pixel level. 

%
%

\subsubsection{Acknowledgements} We thank Will Connell, Sina Ghandian, Parker Grosjean, and Dr. Rodrigo Torres for feedback on the manuscript. This work was supported by grant no. 2018-191905 from the Chan Zuckerberg Initiative DAF, an advised fund of the Silicon Valley Community Foundation (M.J.K.), Department of Defense grant W81XWH2110982, and VA grant 1I01HX003473 (M.L.W.). 

%
%
%
\bibliographystyle{splncs04}

\bibliography{BibTex/library.bib}

\end{document}